\documentclass[11pt]{article}

\usepackage[a4paper,margin=1in]{geometry}
\usepackage{times}
\usepackage{graphicx}
\usepackage{hyperref}
\usepackage{enumitem}
\usepackage{titlesec}
\usepackage{amsmath}
\usepackage{xcolor}

\usepackage[numbers,sort&compress]{natbib}

\begin{document}

\begin{center}

{\Large \bf The unique capabilities of HST for stellar physics}\\[0.7em]
{\bf Probing Atmospheric Structure, Chromospheres, and Mass Loss of Evolved Stars}\\[0.7em]

{\normalsize
Maryam Saberi$^{1}$, 
Graham Harper$^{2}$,
Jacco Th. van Loon$^{3}$, 
Andrea K. Dupree$^{4}$,
Wouter Vlemmings$^{5}$,
Susanne H\"ofner$^{6}$, 
Theo Khouri$^{5}$,
Sven Wedemeyer$^{1}$, 
Roberta Humphreys$^{7}$,
Donald Luttermoser$^{8}$,
Atefeh Javadi$^{9}$,
Pierre Kervella$^{10}$,
Joachim Wiegert$^{6}$
}

\vspace{0.2cm}

{\small
$^{1}$Institute of Theoretical Astrophysics, University of Oslo, Norway;
$^{2}$Center for Astrophysics and Space Astronomy, University of Colorado Boulder, USA;
$^{3}$Lennard-Jones Laboratories, Keele University, UK;
$^{4}$Center for Astrophysics, Harvard \& Smithsonian, USA;
$^{5}$Department of Physics and Astronomy, Chalmers 
University of Technology, Sweden;
$^{6}$Department of Physics and Astronomy, Uppsala University, Sweden;
$^{7}$College of Science and Engineering, University of Minnesota, USA;
$^{8}$East Tennessee State University, USA;
$^{9}$Institute For Research In Fundamental Sciences, IRAN;
$^{10}$Paris Observatory, France

}

\end{center}

\vspace{0.2cm}

\section*{Summary}

Evolved stars are among the primary sources of chemical enrichment and dust production in galaxies. During the giant phases, stars return a substantial fraction of their mass to the interstellar medium (ISM) through stellar winds, enriching galaxies with newly synthesized elements and dust \citep[][]{Goldman2017,HabingOlofsson2003,HoefnerOlofsson2018,Matthews2024,vanLoon2025}. However, the atmospheric structure and physical processes that initiate mass loss remain poorly constrained observationally \citep[][]{FreytagHoefner2023,HoefnerOlofsson2018,vanLoon2025}.

Understanding the origin, structure, and evolution of stellar chromospheres remains a long-standing problem in stellar astrophysics. While the mechanisms responsible for chromospheric heating and atmospheric dynamics are not fully understood even in the Sun, they become more complex in evolved stars due to pulsation, shocks, convection, extended atmospheres, and possible magnetic activity \citep{Airapetian2000, Auriere2010,Harper2001,JudgeStencel1991,Linsky2017,Luttermoser1989,OGorman2020}.

From the main sequence through the red giant branch (RGB), asymptotic giant branch (AGB), red supergiant (RSG), and yellow hypergiant (YHG) phases, the relative roles of pulsation-driven shocks, magnetic activity, and persistent chromospheric heating in producing ultraviolet (UV) emission remain uncertain \citep{Airapetian2000,GordonHumphreys2019,Harper2001,JohnsonLuttermoser1987,Jones2025,Linsky2017,Luttermoser1989,Wood2024}. Determining the thermal, density, and velocity structure of these extended atmospheres is therefore essential for understanding atmospheric heating, the onset of mass loss, and the late stages of stellar evolution \citep{HoefnerOlofsson2018,vanLoon2025,Wood2024}.

High-resolution NUV and FUV spectroscopy ($R\sim$30,000--100,000) provided by HST/STIS occupies a unique observational parameter space that cannot be replaced by existing facilities. JWST lacks UV access, and its infrared detectors can saturate on nearby benchmark AGB and RSG stars, while Roman is not optimized for high-resolution stellar spectroscopy. Ground-based facilities cannot access the UV diagnostics required to probe chromospheres, shocks, ionization structure, and hot gas in stellar atmospheres because of atmospheric absorption. HST/STIS therefore remains essential for understanding the atmospheric physics and mass-loss processes of evolved stars into the 2030s.

We highlight the need to preserve and prioritize high-resolution NUV and FUV spectroscopic capabilities with HST, including coordinated long-term observing programs that establish a legacy dataset of evolved stars. Such programs would provide essential benchmarks for stellar atmosphere modeling, complement ongoing ALMA and optical observations, and help define future UV-optical capabilities for the Habitable Worlds Observatory (HWO) \citep{Linsky2025UVWhitePaper}.


\section{Chromospheres and Atmospheric Structure in Evolved Stars}

While chromospheres in solar-type stars have been extensively studied, it is not known how chromospheres evolve from the main sequence to giant phases. Evolved stars possess extended extended and dynamic atmospheres shaped by pulsation, convection, shocks, and possibly magnetic fields \citep{Auriere2010,Harper2001,JudgeStencel1991,Linsky2017,Wood2024}. The interplay between these processes remains poorly constrained observationally, particularly in the UV, where the key diagnostics of hot gas originate \citep{Carpenter1988,JohnsonLuttermoser1987,Luttermoser1989,Ortiz2019}


Mass loss is now thought to arise from the interplay of pulsation, convection, shocks, atmospheric levitation, dust formation, radiative acceleration, and possibly magnetic or chromospheric heating, rather than from a single mechanism acting in isolation \citep{Airapetian2000,Auriere2010,Bowen1988,BowenWillson1991,FreytagHoefner2023,HoefnerOlofsson2018}. Pulsation and convective motions can lift material above the photosphere, while shocks and chromospheric heating may deposit additional energy in the extended atmosphere; at larger radii, lower temperatures permit molecule and dust formation, and radiation pressure on these grains may then contribute to driving the outflow \citep{Hoefner2008,HoefnerOlofsson2018,FreytagHoefner2023}.

However, recent high-angular-resolution observations from ALMA and optical interferometers reveal that the wind-launching region is highly asymmetric, time-variable, and dynamically structured \citep{Decin2017,Gottlieb2022,Khouri2016,Khouri2019,OGorman2017,Paladini2018,Vlemmings2017}. These observations show that current dust-driven models alone do not reproduce the full range of observed kinematics, asymmetries, and chemistry \citep{FreytagHoefner2023,Gottlieb2022}. The atmospheres of evolved stars appear far more complex than predicted by static or purely dust-driven scenarios \citep{FreytagHoefner2023,vanLoon2025}.

A key missing component in current observational constraints and theoretical models is the hot gas associated with shocks and chromospheric heating, which is primarily accessible through UV diagnostics from abundant species in their dominant ionization states \citep{JohnsonLuttermoser1987,Linsky2017,Luttermoser1989,Wood2024}.


\section{Ultraviolet Diagnostics of Chromospheres and Shocks}

One of the major unresolved questions in stellar physics is the nature of chromospheres in evolved stars, including their heating mechanisms, time variability, and role in mass loss \citep{HoefnerOlofsson2018,JudgeStencel1991,Linsky2017,Wood2024}. While chromospheres have been extensively studied in main-sequence stars and red giants, only a small number of studies have explored them in AGB stars because of the scarcity of UV observations \citep{JohnsonLuttermoser1987,Luttermoser1989,Luttermoser2009,Ortiz2019}.

Early assumptions suggested that the low effective temperatures of AGB stars ($\sim2500$--3000 K) would preclude strong UV emission. However, observations from IUE and HST demonstrated that evolved stars exhibit prominent UV emission lines including Mg~I, Mg~II, C~II], Fe~I, Fe~II, Al~II], and Si~II \citep{Carpenter1988,JohnsonLuttermoser1987,Luttermoser1989,Luttermoser2009,Ortiz2019}. These lines indicate the presence of chromospheric or shock-heated gas in the atmospheres of AGB stars. 

However, the physical origin of this emission remains uncertain. It is still unclear whether the observed UV lines are primarily produced by pulsation-driven shocks, magnetic activity, or persistent chromospheric heating \citep{JudgeStencel1991,Luttermoser1989,Ortiz2019}. Distinguishing between these possibilities requires time-resolved, high-resolution UV spectroscopy capable of tracing atmospheric dynamics throughout the pulsation cycle. 

Phase-dependent velocity variations further demonstrate that the outer atmospheres are dominated by dynamic processes rather than hydrostatic equilibrium \citep{Bowen1988,BowenWillson1991,Wood2000}. Pulsation-driven shocks propagate through the atmosphere and can generate extended regions of enhanced temperature. Hydrodynamic models predict that these warm shocked regions persist throughout much of the pulsation cycle and may form what has been described as a ``hydrodynamic chromosphere'' or ``calorisphere'' \citep{vanLoon2025,Willson2006}.

Despite decades of theoretical work, the physical structure and variability of these regions remain poorly constrained because high-quality UV spectroscopy is extremely scarce.
Fundamental questions therefore remain unanswered:

\begin{itemize}[leftmargin=1.5cm]
\item What is the structure of the hydrodynamic chromosphere in evolved stars?
\item How does the chromosphere evolve throughout the pulsation cycle?
\item What role do shocks and chromospheric heating play in driving mass loss?
\item How does hot shocked gas influence chemistry and dust formation?
\item How do chromospheric properties differ between RGB, AGB, RSG, and YHG stars?
\end{itemize}

High-resolution UV spectroscopy uniquely probes different atmospheric layers through diagnostics formed over a broad range of temperatures and optical depths. Simultaneous observations of neutral and ionized species, including Mg~I/Mg~II, Fe~I/Fe~II, and Si~I/Si~II, provide direct constraints on atmospheric stratification, ionization balance, turbulence, and velocity structure \citep{Carpenter1988,Luttermoser2009,Wood2024}. These diagnostics are essential for constructing realistic atmospheric models of evolved stars and identifying the physical conditions under which mass loss is initiated. 


\section{The Unique Role of HST/STIS}

HST/STIS offers unmatched NUV/FUV spectroscopic capabilities, with no planned replacement expected before HWO in the late 2030s or 2040s.

\subsection{Ultraviolet Access}

Critical diagnostics of chromospheres and shocked gas, including Mg~II h \& k, C~II], Fe~II, and Si~II, are accessible only at UV wavelengths. These lines probe temperature structure, ionization balance, electron density, turbulence, and shock propagation in the upper atmosphere \citep{Linsky2017,Luttermoser2009}. 

\subsection{High Spectral Resolution}

The atmosphere of evolved stars exhibit rich and crowded spectra with strong blending from atomic and molecular transitions \citep{Carpenter1988,Luttermoser2009,Ortiz2019}. Moderate-resolution spectroscopy is insufficient to separate velocity components, shocks, turbulence, and ionization structure. High spectral resolution of STIS ($R\sim$30,000--100,000, corresponding to $\sim10$--3 km s$^{-1}$) is therefore essential \citep{Linsky2025UVWhitePaper,Saberi2026}.

In STIS NUV spectra of the nearby AGB star R~Leo, more than 200 emission lines have been identified \citep{Luttermoser2009}, demonstrating the diagnostic power of high-resolution UV spectroscopy. However, R~Leo remains one of the very few AGB stars with high-quality, broad-wavelength UV spectroscopic observations, highlighting the severe lack of comparable data for evolved stars. $\alpha$~Her~A, observed with high-resolution FUV spectroscopy in HST Cycle 33, similarly reveals dozens of lines tracing hot gas in the outer atmosphere.

The C~II] multiplet lines provide sensitive electron-density diagnostics \citep{Stencel1981}. STIS observations of R~Leo indicate electron densities of order $\sim10^9$~cm$^{-3}$ \citep{Saberi2026}, several orders of magnitude larger than predicted by current dynamical models. This discrepancy strongly suggests that important physical processes remain absent from current theoretical descriptions of evolved stellar atmospheres \citep{BojnordiArbab2024}.

The simultaneous coverage of multiple ionization states and transitions formed at different atmospheric depths allows high-resolution UV spectroscopy to constrain atmospheric stratification in ways that are not possible at other wavelengths. These observations provide critical information about the thermal structure, velocity field, turbulence, and ionization balance in the region where mass loss is initiated. 
Additionally, binarity may contribute to some of the observed asymmetries, heating signatures, and complex mass-loss structures in evolved stars \citep{DeMarcoIzzard2017}. HST's high spatial resolution, through both imaging and spatially resolved STIS spectroscopy, can help identify and separate companion contributions in systems that host binaries.

\subsection{Dynamic Range, Spectral Coverage, and Benchmark Targets}

Nearby evolved stars such as Betelgeuse, R~Leo, and VY~CMa are benchmark objects for stellar-atmosphere and mass-loss studies because they can be spatially resolved and observed across multiple wavelengths \citep{GillilandDupree1996,Vlemmings2017,Humphreys2019,Humphreys2024,Humphreys2025}. However, these nearby stars are often too bright for efficient JWST spectroscopy and can saturate rapidly in several observing modes. In addition, JWST lacks access to the UV diagnostics required to probe chromospheres, shocks, and ionized gas in the upper atmosphere.

For more distant evolved stars, the moderate spectral resolution available in many infrared observing modes can also lead to severe line blending in the highly crowded spectra of their atmospheres, limiting the ability to disentangle velocity fields, ionization structure, and atmospheric stratification. Roman, while transformational for survey science, is not optimized for high-resolution stellar spectroscopy.

HST/STIS therefore occupies a unique observational parameter space that is only planned to be fulfilled again in more than a decade by the HWO.

\subsection{Complementarity with ALMA and JWST}

Although HST/STIS provides unique UV capabilities, the future of evolved-star astrophysics relies on coordinated multi-wavelength observations across several facilities. These observatories probe complementary regions of the stellar atmosphere and circumstellar environment: 

\begin{itemize}[leftmargin=1.5cm]
\item ALMA probes the cool molecular envelope and outer atmosphere, 
\item JWST probes dust formation and infrared molecular chemistry in targets accessible to its dynamic range and spectral capabilities,
\item HST probes chromospheres, shocks, ionization structure, and hot gas in the inner atmosphere,
\item Ground-based optical spectroscopy probes the photosphere and lower atmosphere, including optical shock tracers, atomic and molecular bands.
\end{itemize}

Together, ground-based optical spectroscopy and HST/STIS provide the key constraints on the atmospheric stratification from the photosphere to the chromosphere, while JWST and ALMA extend this picture outward to the dust-formation zone and molecular envelope. 

HST/STIS currently provides the only operational facility capable of obtaining high-resolution UV spectra of evolved stellar atmospheres. The remaining operational lifetime of HST therefore represents a unique and time-critical opportunity to establish a legacy UV dataset before this capability disappears.


\section{The Need for a Community Legacy Program}

Although IUE observations demonstrated phase-dependent UV variability decades ago, only one evolved star, Betelgeuse, has been monitored from 1988 to 2025 \citep{Dupree2026}. A very small number of evolved stars possess modern moderate- or high-resolution UV spectra suitable for detailed analysis. This scarcity of observations now represents a major limitation for the field. Recent ALMA observations and modern hydrodynamic models have transformed our understanding of evolved stellar atmospheres,  but the UV observations required to characterize the hot shocked gas component remain largely absent.

The community would therefore strongly benefit from a coordinated HST legacy program dedicated to high-resolution UV spectroscopy of evolved stars. Such a program should include phase-resolved and long-term STIS monitoring and systematic coverage of RGB, AGB, RSG, and YHG stars spanning different physical properties, and coordinated observations with ALMA and high-resolution ground-based optical spectrographs. These observations would provide critical constraints for NLTE radiative-transfer, chromospheric-heating, and hydrodynamic atmosphere models, as well as for non-equilibrium chemistry, dust formation, mass-loss prescriptions, stellar-evolution models, and galactic chemical-enrichment studies \citep{FreytagHoefner2023,Linsky2017,HoefnerOlofsson2018}. The resulting datasets would remain scientifically valuable for decades and would establish benchmark observations for future UV missions.


\section{Preparing for the Habitable Worlds Observatory}

HST provides a critical bridge toward future UV-optical flagship missions such as the Habitable Worlds Observatory (HWO). The scientific importance of UV spectroscopy extends far beyond exoplanets and includes stellar atmospheres, chromospheric physics, mass loss, shocks, plasma physics, and the baryonic lifecycle within galaxies. Evolved stars provide a particularly compelling demonstration of why UV access and high spectral resolution remain essential capabilities for future observatories. 

Future UV-optical missions will require benchmark science cases and archival datasets to define instrumental requirements and scientific priorities. Evolved stars provide a particularly powerful demonstration of the importance of UV spectroscopy for stellar physics, time-domain astrophysics, and plasma processes in dynamic environments. HST observations obtained during the coming decade would therefore provide both scientific foundations and observational benchmarks for future UV missions.

Preserving HST UV capabilities into the 2030s would therefore:
\begin{itemize}[leftmargin=1.5cm]
\item maintain scientific continuity and establish archival benchmarks for future observatories,
\item refine theoretical models of evolved stellar atmospheres, mass loss, and chemical enrichment,
\item define the UV spectroscopic requirements for next-generation UV-optical facilities.
\end{itemize}

A continued STIS program on evolved stars would therefore not only address major unsolved problems in stellar physics, but also help establish the scientific and technical foundation for future UV missions.


\section{Conclusions}

Understanding the structure, heating, and evolution of chromospheres in evolved stars remains one of the central unsolved problems in stellar astrophysics. At the same time, understanding how mass loss is initiated in evolved stellar atmospheres is critical for explaining the chemical enrichment and dust evolution of galaxies. Recent advances from ALMA and theoretical modeling have revealed the need for high-resolution UV observations capable of probing shocks, chromospheres, atmospheric stratification, and ionized gas in the inner atmospheres of evolved stars. HST/STIS uniquely provides this capability. No existing or planned facility offers the combination of UV access, high spectral resolution, sensitivity, and dynamic range needed to address these questions. As Hubble enters the 2030s, preserving and prioritizing its UV spectroscopic capabilities will enable transformational progress in stellar astrophysics, provide essential synergy with JWST and ALMA, and help define the scientific foundations for future UV-optical missions.

\end{document}